\documentclass[10pt, a4paper, compsoc, conference]{IEEEtran}
\ifCLASSINFOpdf
  \usepackage[pdftex]{graphicx}
  \graphicspath{{../pdf/}{../jpeg/}}
  \DeclareGraphicsExtensions{.pdf,.jpeg,.png}
\else
\fi
\usepackage{listings}
\usepackage{color}

\usepackage{xcolor}
\definecolor{backcolour}{rgb}{1,1,1}
\colorlet{punct}{red!60!black}
\definecolor{delim}{RGB}{20,105,176}
\colorlet{numb}{magenta!60!black}

\lstdefinelanguage{json}{
    basicstyle=\ttfamily\footnotesize,
    numberstyle=\scriptsize,
    stepnumber=1,
    numbersep=5pt,
    showstringspaces=false,
    breaklines=true,
    backgroundcolor=\color{backcolour},
    literate=
     *{0}{{{\color{numb}0}}}{1}
      {1}{{{\color{numb}1}}}{1}
      {2}{{{\color{numb}2}}}{1}
      {3}{{{\color{numb}3}}}{1}
      {4}{{{\color{numb}4}}}{1}
      {5}{{{\color{numb}5}}}{1}
      {6}{{{\color{numb}6}}}{1}
      {7}{{{\color{numb}7}}}{1}
      {8}{{{\color{numb}8}}}{1}
      {9}{{{\color{numb}9}}}{1}
      {:}{{{\color{punct}{:}}}}{1}
      {,}{{{\color{punct}{,}}}}{1}
      {\{}{{{\color{delim}{\{}}}}{1}
      {\}}{{{\color{delim}{\}}}}}{1}
      {[}{{{\color{delim}{[}}}}{1}
      {]}{{{\color{delim}{]}}}}{1},
}

\usepackage{multirow}
\usepackage{hhline}

\usepackage[inline]{enumitem}

\begin{document}
\title{A web service based on RESTful API and JSON Schema/JSON Meta Schema to construct knowledge graphs}

\author{
\IEEEauthorblockN{Adam Agocs}
\IEEEauthorblockA{CERN\\
CH-1211 Geneva 23\\
Switzerland\\
Email: adam.agocs@cern.ch}
\and
\IEEEauthorblockN{Jean-Marie Le Goff}
\IEEEauthorblockA{CERN\\
CH-1211 Geneva 23\\
Switzerland\\
Email: jean-marie.le.goff@cern.ch}
}
% \author{Adam Agocs\thanks{CERN, Geneva 23, Geneva, Switzerland},
%         Jean-Marie Le Goff
% \thanks{M. Shell was with the Department
% of Electrical and Computer Engineering, Georgia Institute of Technology, Atlanta,
% GA, 30332 USA e-mail: (see http://www.michaelshell.org/contact.html).}% <-this % stops a space
% \thanks{J. Doe and J. Doe are with Anonymous University.}% <-this % stops a space
% \thanks{Manuscript received April 19, 2005; revised August 26, 2015.}}


\markboth{Journal of \LaTeX\ Class Files,~Vol.~14, No.~8, August~2015}%
{Shell \MakeLowercase{\textit{et al.}}: Bare Demo of IEEEtran.cls for IEEE Journals}



\maketitle

\begin{abstract}
Data visualisation assists domain experts in understanding their data and helps them make critical decisions. Enhancing their cognitive insight essentially relies on the capability of combining domain-specific semantic information with concepts extracted out of the data and visualizing the resulting networks. Data scientists have the challenge of providing tools able to handle the overall network lifecycle. In this paper, we present how the combination of two powerful technologies namely the REST architecture style and JSON Schema/JSON Meta Schema enable data scientists to use a RESTful web service that permits the construction of knowledge graphs, one of the preferred representations of large and semantically rich networks. 
\end{abstract}


% \fi
\IEEEpeerreviewmaketitle

\section{Introduction}\label{section_intro}
\IEEEPARstart{C}{ollaboration Spotting (CS)} \cite{ERCIM_NEWS:2107}\cite{collspotting} is a platform designed to support visual analytics of multivariate knowledge graphs built out of data from heterogeneous  sources. It offers a novel approach to handle semantic and structural complexity at the interactive visualisation level by enabling users to generate different perspectives of domain-related knowledge graphs, navigate between these perspectives and execute different graph algorithms within them. 

Since 2012, CS has been deployed on many pilot projects in various domain-related analysis to demonstrate its capability of enhancing the cognitive insight of humans into the understanding of their data. In particular, CS has been used
\begin{itemize}
\item to analyse publications and patents for dental science \cite{dentistry:paper} and for the detection of technology and innovation developments \cite{Joanny2015MonitoringOT},  
\item for a security-threat analysis \cite{webpage:UNICRI}, 
\item for a university ranking project in collaboration with the MTA-PE Budapest Ranking Research Group \cite{webpage:Telcs} and 
\item for a neuroscience project in collaboration with the Complex Systems and Computational Neuroscience Group \cite{webpage:Negyessy} at Wigner RCP. 
\end{itemize}

CS capability of enhancing cognitive insight strongly depends on the construction of domain-independent and semantically rich knowledge graphs. 
In essence, constructing such graphs calls for an API that enables users to:
\begin{enumerate}
\item use descriptors to specify data elements with minimal restrictions, \label{req:first}
\item structure data elements in an ontology-like hierarchy, \label{req:two} 
\item set up data-related control parameters to satisfy the data-driven approach of the platform, \label{req:third}
\item validate descriptors by using pre-defined node and edge descriptors, \label{req:fourth}
\item validate data with their descriptors and \label{req:fifth}
\item upload data and  descriptors by using well-known and widespread technologies. \label{req:sixth}
\end{enumerate}
To these, one must add requirements that are specific to the interactive visualization and navigation aspects of the CS platform: 
\begin{enumerate}
\setcounter{enumi}{6}
\item Use of a graph database (which is a natural choice in support of visual analytics on graphs) such as Neo4j \cite{neo4j:manual} for storing \begin{enumerate*}[label=\textbf{\alph*)}] \item users' data as knowledge graphs i.e. a 4-element tuples ($G=(V, E, L, \alpha)$) where both vertices and edges are labelled and \item the knowledge graph's schema called as reachability graph (please, see \cite{agocs2017interactive} for further details).\end{enumerate*}\label{req:seventh}
\item Use of Django \cite{manual:djangoWeb} as a web framework implemented in Python. \label{req:eighth}
\item Use of JSON \cite{ECMA:JSON}\cite{rfc:JSON_interchange} as a data exchange format between the different layers of the platform. \label{req:ninth}
\end{enumerate}
A REST architecture like web service combined with JSON Schema/JSON Meta Schema \cite{json_schema:webpage}\cite{ietf:JSON_schema} and Py2neo Python package \cite{python:py2neo} provides an adequate tool for constructing knowledge graphs in compliance with the above-mentioned requirements.

After the related work emphasising JSON format and enabling technologies, Section \ref{section_architecture} presents the architecture of the RESTful web service of the Collaboration Spotting platforms showing how the selected technologies satisfy the requirements and how an extension of the JSON Schema specifications can support an ontology-like hierarchy for the descriptors. Section \ref{section_use_case} gives a use-case and some experimental results on the scalability of the API and some comparisons between the single mode and the bulk mode operations. And finally, the paper finishes with Future Work and Conclusion in Section V and Section VI respectively.

\section{Related Work}\label{section_related_work}
In his PhD dissertation, Roy Fielding described the principled design of the modern web architecture that leads to the REST architecture style \cite{REST:first:fielding2000architectural}. Since then, REST gained in popularity amongst the API (Application Programming Interface) developers and became the most used approach for developing web services \cite{programmableWeb:pres:vitvar2010programmableweb}\cite{programmableWeb:the_website}. According to programmableWeb.com \cite{programmableWeb:the_website}, most of these services have been developed with API using the JSON (JavaScript Object Notation) format to send and receive requests and responses over the HTTP protocol. Since its draft submission, JSON has followed an XML-like path starting as a data exchange format over the Internet to become part of an exchange protocol used by various APIs. The most notable ones being:
\begin{itemize}
\item JSON-LD \cite{JSON_LD:Lanthaler:2012:UJC:2307819.2307827} (W3C recommendation \cite{w3c:json-ld}), a JSON-based serialisation for handling Linked Data \cite{bizer2009linked} extended with contextual explanations,
\item JSON API \cite{json:api} specifies the communication protocol between clients and servers,
\item JSON-RPC \cite{json:rpc} a remote procedure calls in JSON, similar to XML-RPC with XML.
\end{itemize}

JSON-to-RDF/XML converters provide an indirect means to valid native JSON documents. To the best of our knowledge, JSON Schema is the only format that enables users to define the syntax of a JSON document. Python supports the fourth draft version of this format, but a more recent version is available (the seventh one). Although JSON schema format is still in the process of standardisation, the number of JSON Schema-based theoretical and practical results is growing. In particular, this is the case for the schema's formal definition \cite{JSON_Schema:first:Pezoa:2016}) and SDMX-JSON data retrieval of OECD \cite{OECD:JSON} datasets.

The JSON key features that are the validation rules (JSON Meta Schema) and the code to validate these rules made it possible to develop node and edge validators for Collaboration Spotting.
Since 2004, OWL \cite{owl} has become a W3C supported standard for query languages on ontologies. OWL and ontologies for domain-specific knowledge representations are key technologies in various research areas such as data retrieval, data mining, machine learning and data visualisation. The criteria of Section \ref{section_intro}, does not require the creation of an OWL-equivalent language built upon JSON schema. Applying the Amann and Fundulaki’s mathematical definition \cite{amann1999integrating} of an ontology combined with a graph database supporting a labelled property graph data model will be sufficient for the construction of knowledge graphs compliant with the above requirements. In Amann and Fundulaki's work, an ontology is expressed as a triplet, $\mathcal{O} = (C, R, isa)$, where
\begin{enumerate*}
\item $C$ is a set of concepts, 
\item $R$ is a set of binary typed roles between these concepts and 
\item $isa$ is a set of inheritance ("is a") relationship between them.
\end{enumerate*}

\section{Architecture}\label{section_architecture}
As indicated above, the CS RESTful API is built upon three main technology elements, namely the Django REST framework, the Python implementation of JSON Schema/JSON Meta Schema and Py2neo Python package for the Neo4j graph database. These technologies address the user requirements of Section \ref{section_intro} as follows:
\begin{itemize}
\item[1)] Descriptors in JSON Schema format can describe user data,
\item[4)] the validation of descriptors in JSON schema requires modifications of the JSON Meta Schema such that each descriptor either describes a graph node or a graph edge in the knowledge graph,
\item[5)] JSON documents containing user data are validated by using descriptors written in the JSON Schema format,
\item[6)] Django REST framework is a powerful and flexible toolkit for building Web APIs that is compliant with the REST architecture style,
\item[7)] Py2neo is a client library for working with Neo4j,
\item[8)] Django REST framework can easily be combined with Django Web framework and
\item[9)] Django REST framework and JSON Schema/JSON Meta Schema support JSON format.
\end{itemize}
Essentially all the user requirements are addressed with the exception of requirements \ref{req:two} and \ref{req:third}.

\subsection{Descriptor}
Requirement \ref{req:two} requires an extension of JSON schema specifications in the form of additional keywords (see Table \ref{json_schema:extra_keywords}) in order to support the correspondence between descriptors and concepts together with the inheritance mechanism as requested by Amann and Fundulakis in their definition of an ontology \cite{amann1999integrating}. More precisely, additional keywords are needed to specify that:
\begin{itemize}
\item a set of concepts (definition of $C$) corresponds to a set of descriptors,
\item edge descriptors can specify binary-typed roles (definition of $R$) between node descriptors (i.e. concepts), and
\item the use of keyword “parents” in node descriptors characterises the inheritance relationship (definition of $isa$), see Example \ref{json_schema:json:he}.
\end{itemize}

\begin{center}
\renewcommand{\lstlistingname}{Example}
\begin{lstlisting}[language=json, caption={JSON Schema for the "High Education" node in the Ranking project \cite{webpage:Telcs}}, captionpos=b, label={json_schema:json:he}]
  {
   "$schema": "http://localhost:8000/schemas/validators/node_validator.json#",
   "id": "http://localhost:8000/schemas/ranking/   he.json#",
   "title" : "HE",
   "type" : "object",
   "properties" : {
    "id": {"$ref": "../basic/basic_definitions.json# /definitions/id"},
    "name" : {"type": "string", "maxLength": 1000,  "minLength": 1}
   },
   "additionalProperties" : {"$ref": "../basic/basic_definitions.json#/definitions/default_property"},
   "required": ["id", "name"],
   "parents" : ["institute"],
   "graph_element" : "node"
  }
\end{lstlisting}
\end{center}

Two restrictions on keyword values apply when writing descriptors:
\begin{itemize}
\item The value of keyword \textit{\$schema} must be either \textit{node\_validator.json} or \textit{edge\_validator.json} and
\item keyword value \textit{required} must contain values \textit{id} and \textit{name}. 
\end{itemize}

\textit{settings} is an additional keyword that is needed to address Requirement \ref{req:third}. \textit{settings} is a list of key-value pairs where each pair makes a connection between a pre-defined function (key) and its attribute (value).
% The seventh requirement is about dealing with Neo4j graph database. This part is implemented by using py2neo v.2. Python package. 

\begin{table*}[!t]
\renewcommand{\arraystretch}{1.3}
\caption{Extra keywords for descriptors}
\label{json_schema:extra_keywords}
\centering
\begin{tabular}{c||c||c||c}
\hline
\bfseries Key & \bfseries Data type & \bfseries Scope & \bfseries Description \\ \hline\hline
graph\_element & "node" $\vert$ "edge" & each descriptor & Mandatory, defines the role of the descriptor\\
parents & array of references & node descriptors & Optional, defines \textit{ISA} connections between node descriptors\\
direction & "double" $\vert$ "single" & edge descriptors & Mandatory, defines whether the edge descriptor is directed or undirected\\
source\_label & reference & edge descriptors & Mandatory, labels the source nodes\\
target\_label & reference & edge descriptors & Mandatory, labels the target nodes\\
\hline
\end{tabular}
\end{table*}

\subsection{Main processes and modules}
\begin{figure}[!tbh]
\centering
 \includegraphics[page=2, height=2.5in]{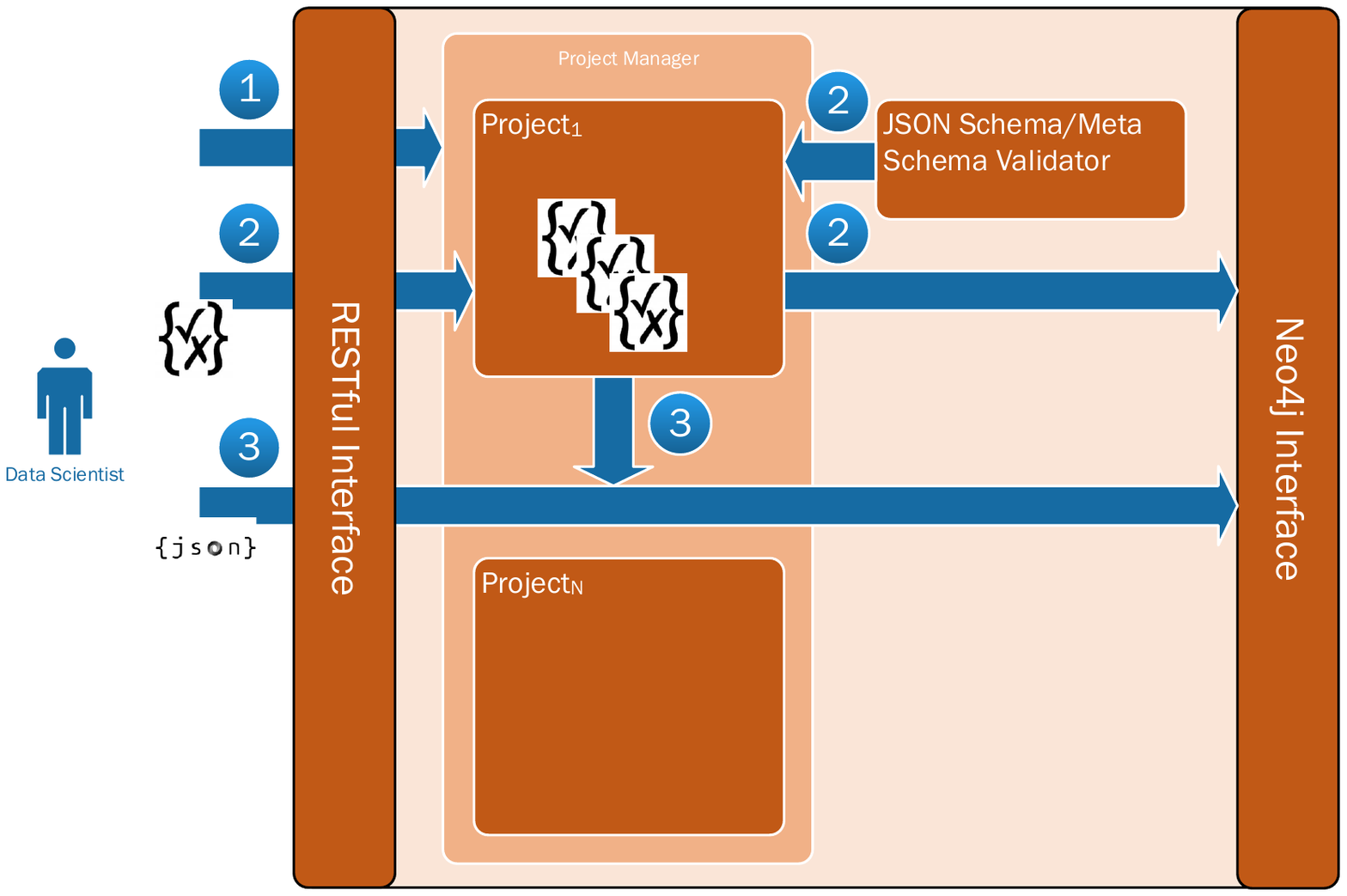}
\caption[1]{Simplified architecture model of Collaboration Spotting's API with the three main processes \begin{enumerate*}\item create projects, \item upload descriptors and \item upload data. \end{enumerate*}} 
\label{figure:RESTful architecture}
\end{figure}

\begin{figure}[!tbh]
\centering
\includegraphics[page=1, height=2.5in]{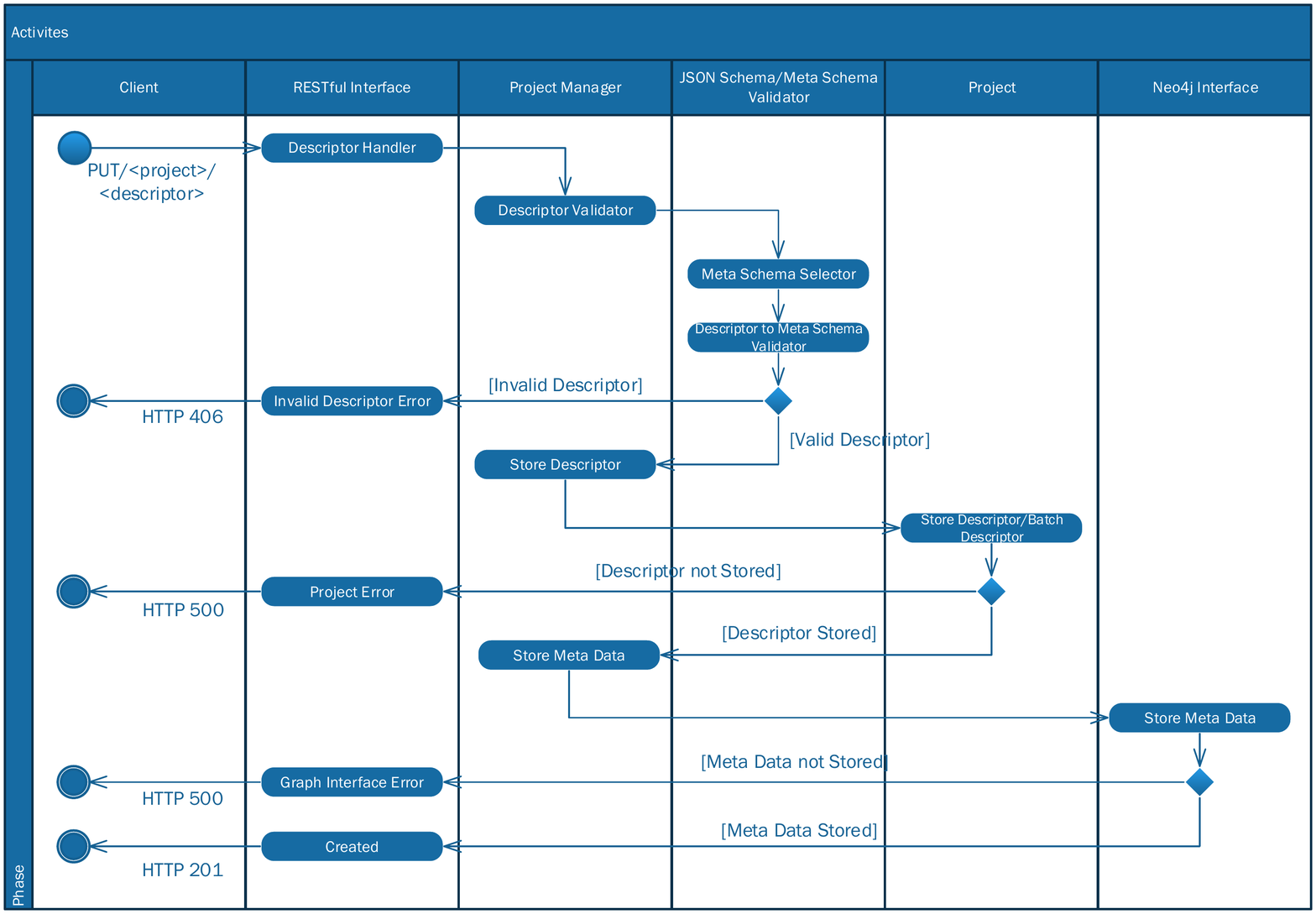}
\caption{UML Activity Diagram of the \textit{upload descriptor} process} 
\label{figure:uploading:descriptor}
\end{figure}

\begin{figure}[!tbh]
\centering
\includegraphics[page=3, height=2.5in]{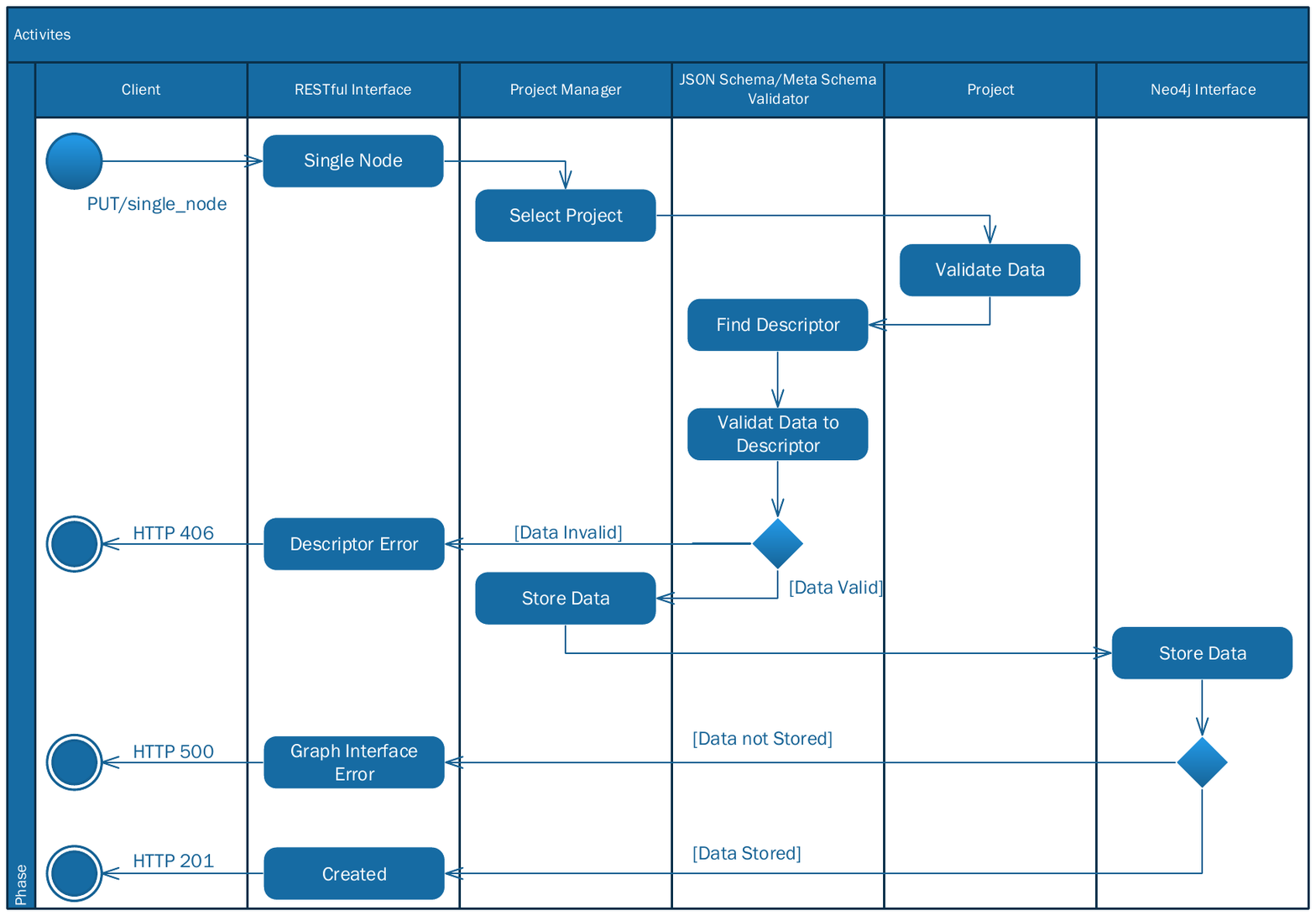}
\caption{UML Activity Diagram of the \textit{upload data} process}
\label{figure:uploading:data}
\end{figure}

Figure \ref{figure:RESTful architecture} shows a simplified architecture model of the CS RESTful API with its three main processes \textit{create project}, \textit{upload descriptors} and \textit{upload data}. These processes make use of the following supporting modules:
\begin{itemize}
\item \textbf{RESTful Interface} uses Django REST Framework. This module is responsible for handling and parsing every RESTful message from clients and sending back response messages,
\item \textbf{Neo4j Interface} maps the RESTful calls to Neo4j’s Cypher queries. 
\item \textbf{JSON Schema/Meta Schema Validator} module validates user descriptors and data in JSON format according to pre-defined descriptors and JSON Meta Schema,
\item \textbf{Project} module generates a bulk descriptor for each user-defined ones and handles single and bulk versions of descriptors. See the “High Education” descriptor of Example \ref{json_schema:json:he} related to the Ranking project and its bulk version in Example \ref{json_schema:json:bulk_he}. Users can validate and upload multiple graph elements in one bulk.
\item \textbf{Project Manager} module supports project creation and selection and forwards messages between the RESTful Interface and a given project.
\end{itemize}
Figure \ref{figure:uploading:descriptor} and \ref{figure:uploading:data} show the UML Activity Diagrams for the upload descriptor and upload data processes respectively. The source code found in Collaboration Spotting GitHub repository \cite{Agocs:code} follows these diagrams.

% The overview of our architecture with the interactions of the three mainly used processes is presented in Figure \ref{figure:RESTful architecture}. It contains the following modules:

% \begin{itemize}
% \item[] \hspace{-\topsep}\textbf{RESTful Interface)} This module implemented by using Django REST Framework, takes responsibility for handling, parsing every RESTful message from the clients and sending the response messages to them.
% \item[] \hspace{-\topsep}\textbf{Neo4j Interface)} This interface does the mapping from RESTful calls to Neo4j's Cypher queries.
% \item[] \hspace{-\topsep}\textbf{JSON Schema/Meta Schema Validator)}  It takes responsibility for validating the uploaded descriptors written in JSON Schema to pre-defined JSON Meta Schemas and user data to the descriptors.
% \item[] \hspace{-\topsep}\textbf{Project)} The task of this module is twofold:
% \begin{itemize}
% \item First, it generates a bulk descriptors for each user-defined descriptor. Example \ref{json_schema:json:he} shows the descriptor of the "High Education" node of our Ranking project, meanwhile, Example \ref{json_schema:json:bulk_he} shows the bulk version of this descriptor which allows the users to validate and upload multiple graph elements as one piece.  
% \item Second, it handles single and bulk version of descriptors together.
% \end{itemize}
% \item[] \hspace{-\topsep}\textbf{Project Manager)} It takes responsibility for managing (create or select) the requested project and forwarding the messages between the RESTful Interface and the given project.
% \end{itemize}

\begin{center}
\centering
\renewcommand{\lstlistingname}{Example}
\begin{lstlisting}[language=json, caption={Automatically generated bulk version of descriptor "High Education" in the Ranking project.}, captionpos=b, label=json_schema:json:bulk_he]
{
 "$schema": "http://localhost:8000/schemas/validators/node_validator.json#",
 "id": "http://localhost:8000/schemas/ranking/bulk_he.json#",
 "title" : "Bulk HE",
 "definitions" : {
  "node" : {"$ref": "./he.json#"}
 },
 "type" : "array",
 "items": {"$ref" : "#/definitions/node"}
}
\end{lstlisting}
\end{center} 

% Note that the IEEE typically puts floats only at the top, even when this
% results in a large percentage of a column being occupied by floats.

\section{Use-case}\label{section_use_case}
\begin{figure}[!tbh]
\centering
\includegraphics[page=4, height=2.5in]{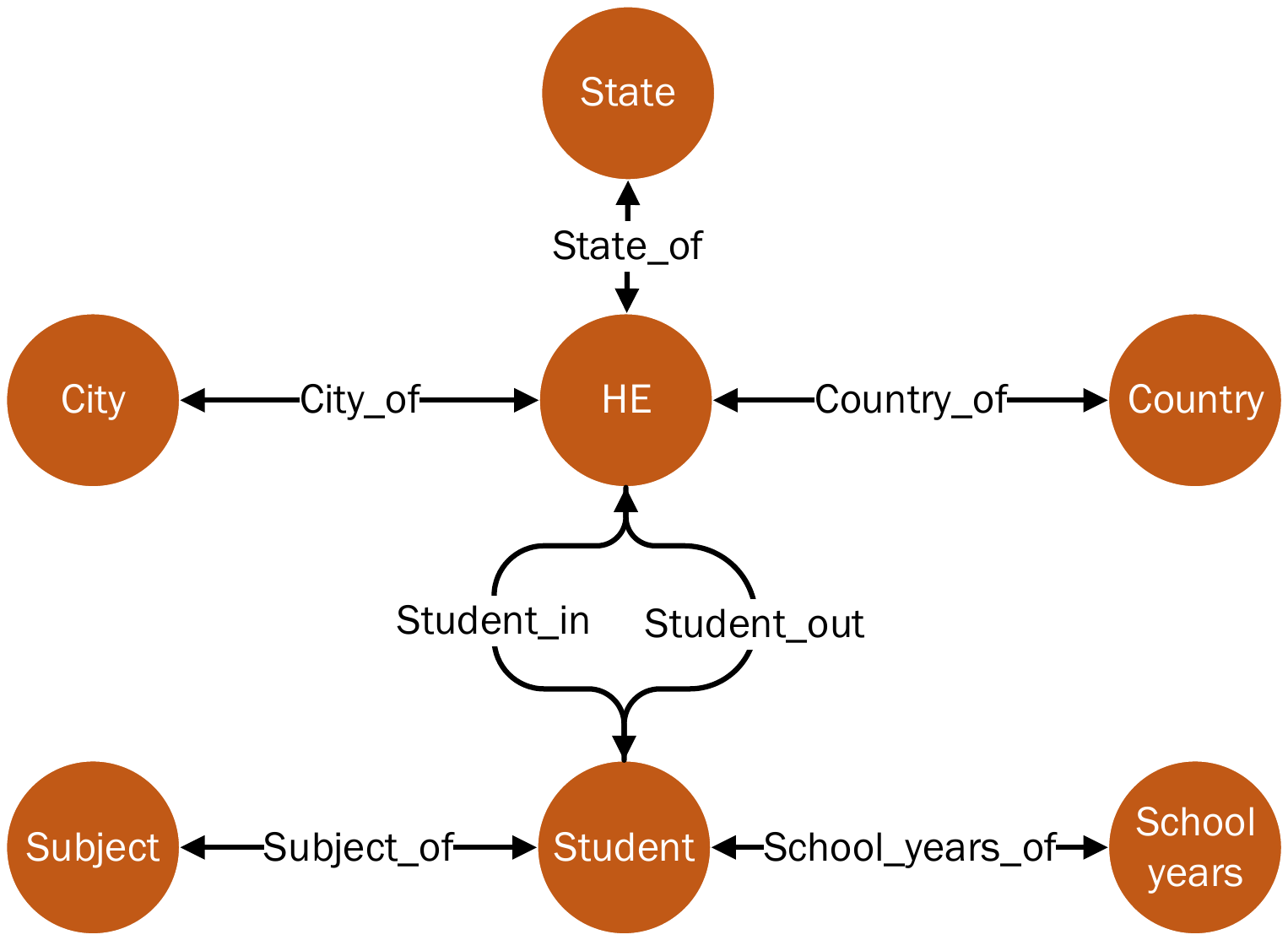}
\caption{The simplified database model for the Ranking project. The $parent$ label of the descriptor in Example \ref{json_schema:json:he} has been omitted.} 
\label{figure:ranking:schema}
\end{figure}

The MTA-PE Ranking Research Group provided a collection of anonymised student travels in the framework of the European Commission’s Erasmus programme during 2008-9 to 2013-14. The Ranking Research Group edited the data in order to merge different time-intervals and selected the relevent attributes for their analysis. A small sample of the results is shown in Table \ref{data:erasmus}. Figure \ref{figure:ranking:schema} gives a simplified data model of the knowledge graph resulting from uploading in Neo4j the data from the Ranking Research Group. One can note that the node of Example \ref{json_schema:json:he}-\ref{json_schema:json:bulk_he} appears in as a node labelled $HE$ with its connecting relationships in the data model.  

 The measurements address the performance of the CS RESTful API when uploading single and bulk inserts. These measurements have been done on an HP Z440 workstation (Intel Xeon E5-1620 v3, 3,50 GHz processor, 32 GB DDR4 RAM and 512 GB SSD drive), using the community version of Neo4j 2.3.1 with -Xmx8192m JVM settings.
 
\begin{figure}[!tbh]
\centering
\includegraphics[width=3.5in]{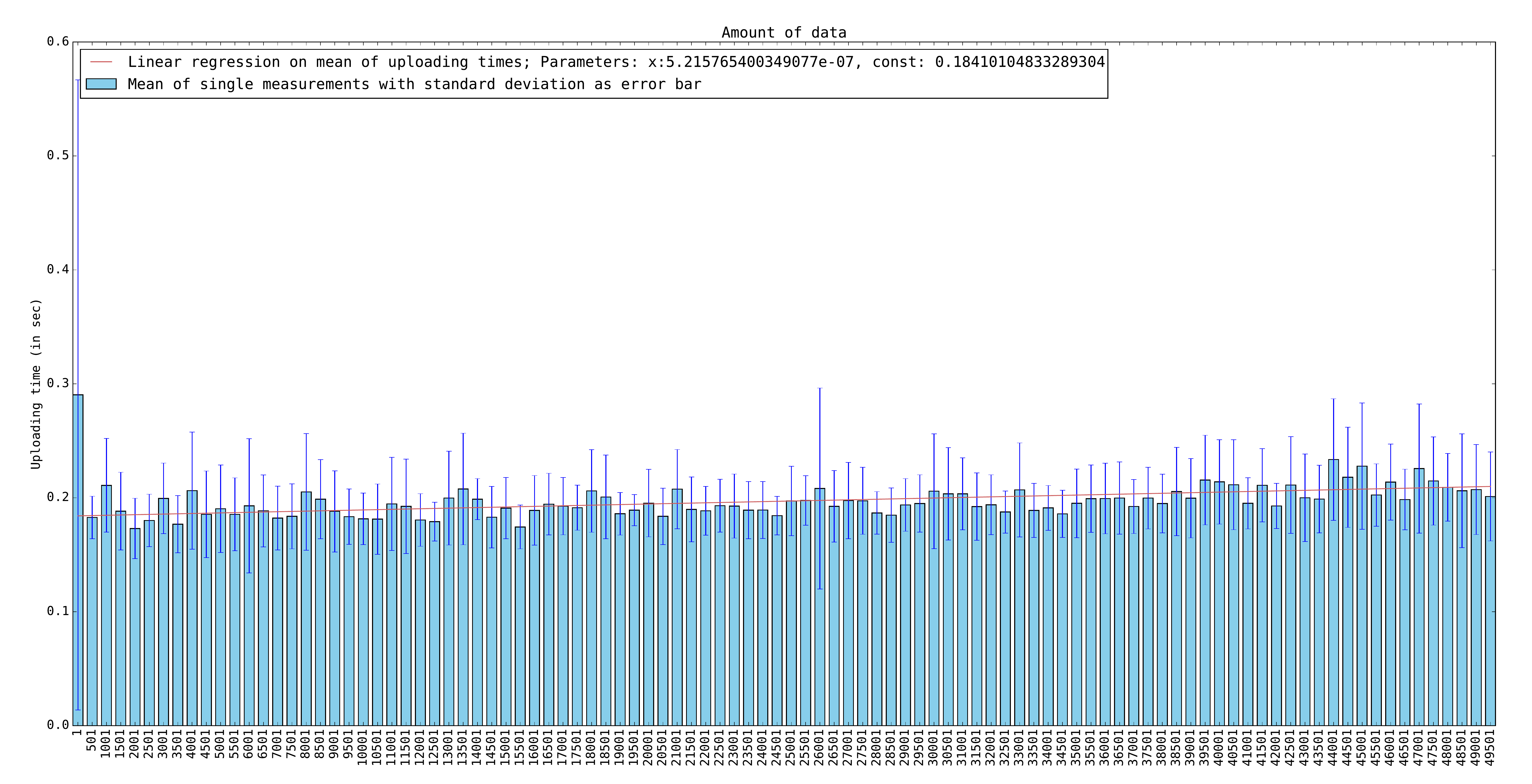}
\caption{Uploading time of single JSON documents} 
\label{figure:ranking:single_measurement}
\end{figure}

\begin{figure}[!tbh]
\centering
\includegraphics[page=1, width=3.5in]{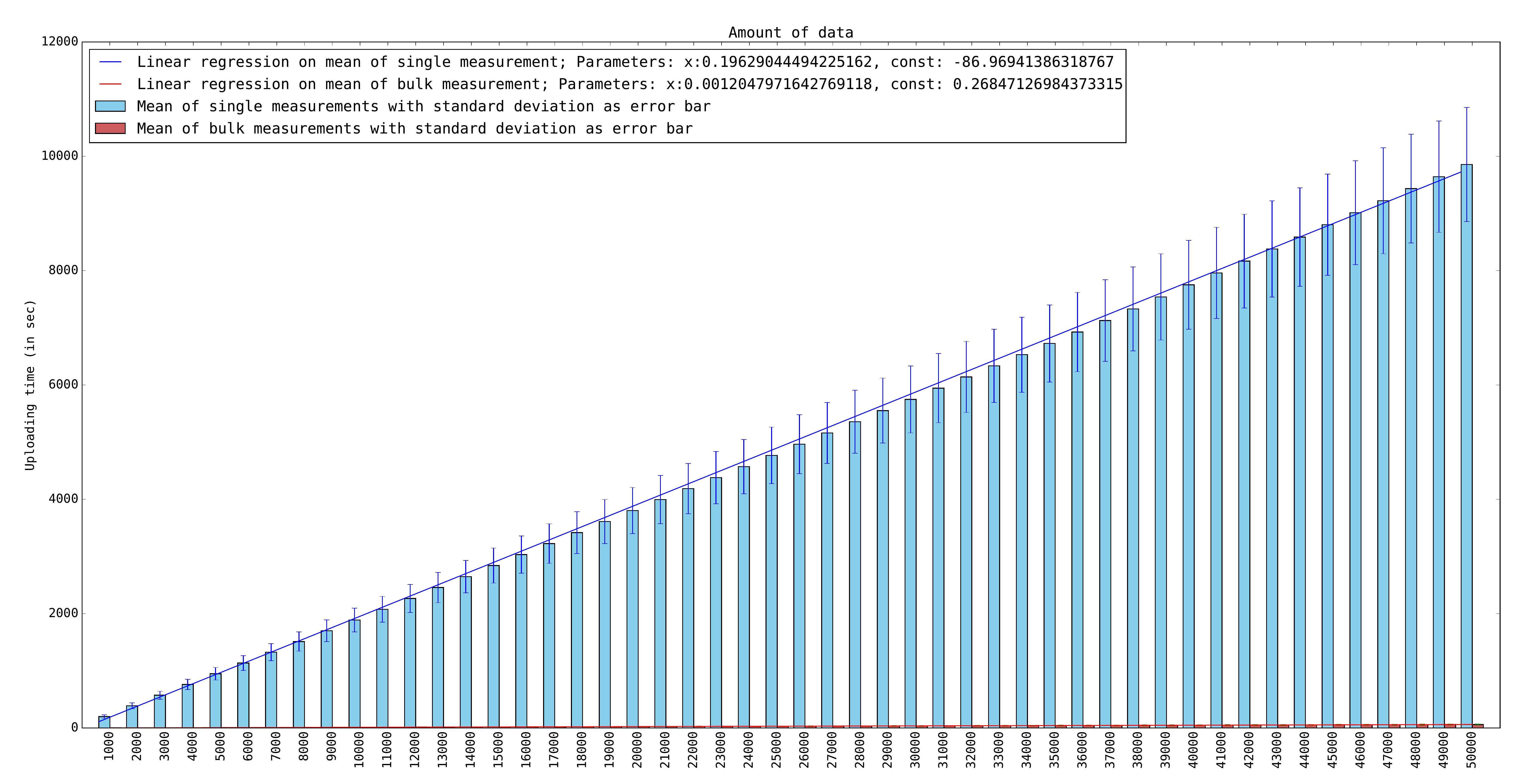}
\caption{Comparison of single and bulk inserts} 
\label{figure:ranking:measurement}
\end{figure}

 Figure \ref{figure:ranking:single_measurement} shows the uploading time of a single JSON document, the descriptor being pre-uploaded. The variations observed depend on the writing time of the JSON document in the Neo4j database. Time, which is directely related to the size of the previously uploaded data in the \textit{upload data} process as can be seen in the activity diagram of Figure \ref{figure:uploading:data}.
Fig. \ref{figure:ranking:measurement} plots the uploading time of single and bulk inserts. It shows that the bulk insert is approximately 163 times faster than the single insert, confirming the clear advantage of using bulk inserts. 
%\begin{itemize}
%\item measuring the uploading time of a single JSON document (the descriptor is pre-uploaded). As you can see in Figure \ref{figure:ranking:single_measurement}, the uploading time is increasing in time. The reason for this effect is the database because it is the only element of the uploading process shown in Figure \ref{figure:uploading:data} which depends on the size of the previously uploaded data.
%\item measuring and comparing the uploading time of single and bulk insert modes. As we can see in Figure \ref{figure:ranking:measurement}, it is worth using bulk insert mode because we reached approximately *.* speed-up.
%\end{itemize}

\begin{table*}[!t]
\renewcommand{\arraystretch}{1.3}
\caption{Erasmus student data}
\label{data:erasmus}
\centering
\begin{tabular}{c||c||c||c||c||c||c||c}
\hline
\bfseries from he & \bfseries from he country & \bfseries to he & \bfseries to he country & \bfseries subject & \bfseries year & \bfseries distance & \bfseries direction \\ \hline\hline 
D  KONSTAN01 & DE & K BATH01 & UK & 3 & 2008-09 & 907 &296,286297877918 \\ 
D  KONSTAN01 & DE & F  PARIS007 & FR & 4 & 2008-09 & 501 &283,942344291399 \\
\vdots & \vdots & \vdots & \vdots & \vdots & \vdots & \vdots & \vdots \\
\hline
\end{tabular}
\end{table*}

\section{Future Work}\label{section_future_work}
As mentioned in the introduction, JSON Schema is not a standard format yet. We will strive to support the new version of the standard as soon as it becomes available. The replacement of simple JSON documents with ones written in JSON-LD could be an alternative solution. It would enable to combine the expressive power of JSON-LD with the verification power of JSON Schema. But the development of the validation code required for JSON-LD would call for a substantial amount of work in comparison with the gain in expressiveness. Indeed, the implementation of the ontology-like hierarchy in the data model is sufficient to construct the knowledge graphs supported by Collaboration Spotting.

\section{Conclusion}\label{section_conclusion}
In this paper, we present the architecture of a RESTful web service based on the combination of the JSON-based REST architecture style with JSON Schema/JSON Meta Schema that permits the construction of knowledge graphs. With this service, users of the Collaboration Spotting platform are able in  rather unique way and with few restrictions, to create the elements of their knowledge graph(s) and validate the uploaded data togheter with these elements by using JSON Schema. A small extension of the JSON Meta Schema, made it possible to extend user-created descriptors to support ontology concepts together with an appropriate validation of these descriptors.

% or



\section*{Acknowledgment}
We would like to thank A. Telcs, Z. T. Kosztyan and L. Gadar from the MTA-PE Budapest Ranking Group for providing the data used in the examples used in this paper and D. Dardanis from the Collaboration Spotting team for his contribution to the visualization of the Ranking knowledge graph.

\ifCLASSOPTIONcaptionsoff
  \newpage
\fi



%\bibliographystyle{IEEEtran}
\bibliographystyle{IEEEtran}
\bibliography{bare_jrnl.bib} 


% \end{IEEEbiography}


%\newpage



%\vfill


\end{document}